**The Adaptive Communication Framework (ACF)**

**for Extraterrestrial Intelligence Discovery**

Omer Eldadi[1], Gershon Tenenbaum[1] and Abraham Loeb[2]

1. B. Ivcher School of Psychology, Reichman University, Herzliya, Israel

2. Department of Astronomy, Harvard University, Cambridge, MA, USA

**Corresponding Address**:

Omer Eldadi

B.Ivcher School of Psychology

Reichman University

The University 8, Herzliya, Israel

Email: Omereldadi@gmail.com

**Abstract**

The Vera C. Rubin Observatory's Legacy Survey of Space and Time will increase interstellar object detection rates to one every few months, substantially elevating the probability of identifying objects with characteristics suggesting artificial origin. Despite this imminent capability, no evidence-based crisis communication framework exists for managing potential technosignature discoveries. We present the Adaptive Communication Framework (ACF), a theoretically grounded protocol that integrates crisis communication theories with the SPECtrum of Rhetoric Intelligences model to address diverse cognitive, social and emotional processing styles. Through analysis of communication failures during COVID-19, Fukushima, and asteroid 99942 Apophis (2004 MN$_4$), we identify critical gaps in managing scientific uncertainty under public scrutiny. The ACF provides graduated protocols calibrated to the Loeb Scale for Interstellar Object Significance, offering specific messaging strategies across four rhetoric intelligence channels (Systematic, Practical, Emotional, Creative) for each evidential level group. Recognizing that Artificial Intelligence (AI) systems will mediate public understanding, the framework incorporates safeguards against synthetic media manipulation and algorithmic misinformation through pre-positioned content seeding and deepfake detection protocols. Our theoretical framework reveals that successful communication of paradigm-shifting discoveries requires simultaneous activation of multiple cognitive channels, with messages adapted to cultural contexts and uncertainty levels. This framework provides essential theoretical groundwork for ensuring humanity's potentially most transformative discovery unfolds through understanding rather than chaos.

*Keywords*: extraterrestrial intelligence, crisis communication, interstellar objects, scientific communication, technosignatures, public engagement

## Introduction

The Vera C. Rubin Observatory's Legacy Survey of Space and Time (LSST) represents a fundamental shift in humanity's ability to detect interstellar objects (ISOs), increasing discovery rates from one per decade to potentially one every few months (Dorsey et al., 2025; Hoover et al., 2022; Siraj & Loeb, 2022). The enhanced detection capability transforms the search for extraterrestrial intelligence from passive listening to active observation of physical objects traversing the solar system. While most ISOs will prove to be natural in origin, the increased sample size raises the statistical probability of detecting objects with characteristics indicative of artificial origin.

The discovery of 1I/'Oumuamua in 2017 revealed both the scientific opportunity and communication challenge posed by anomalous ISO. Its extreme shape and unexpected non-gravitational acceleration without visible outgassing (Bialy & Loeb, 2018) generated intense scientific debate and public speculation, revealing the absence of established protocols for communicating potential technosignature discoveries. Subsequently, the scientific community was challenged by a critical gap, e.g., despite decades of SETI protocols for radio signals (Carrigan, 2006; Gertz, 2016; Lemarchand & Tarter, 1994), to our knowledge, a solid and practical concept for communicating discoveries of potentially artificial physical objects is lacking.

Public readiness for such discoveries has reached an unprecedented level. Recent surveys indicate that 65% of Americans believe that intelligent extraterrestrial life likely exists (Pew Research Center, 2021), while 58.2% of astrobiologists consider extraterrestrial intelligence probable (Vickers et al., 2025). The convergence of enhanced detection capability, public expectation, and absent communication protocols creates conditions for potential social disruption. The recently proposed *Loeb Scale* provides a classification system for ISO anomalies from 0 (*clearly natural*) to 10 (*confirmed technological*) but lacks accompanying communication strategies calibrated to each evidential level (Loeb, 2025a).

Here, we present the development of Adaptive Communication Framework (ACF), the first theoretically grounded protocol specifically designed for communicating ISO-related discoveries that may indicate extraterrestrial intelligence. By integrating established crisis communication theories with the SPECtrum of Rhetoric Intelligences model (Teer, 2020) and lessons from analogous communication challenges, the ACF addresses the unique challenges of managing scientific uncertainty, public psychology, and global coordination during humanity's potentially most transformative discovery.

## Theoretical Foundations of Crisis Communication

### Core Crisis Communication Models

Modern crisis communication theory recognizes that effective emergency messaging must simultaneously convey accurate information, maintain public trust, prevent panic, and enable appropriate response behaviors, specifically these communication processes are designed to reduce and contain harm, provide specific information to stakeholders, initiate and enhance recovery, manage image and perceptions of blame and responsibility, repair legitimacy, generate support and assistance, explain and justify actions, apologize, and promote healing, learning, and change (Seeger et al., 2003). The Crisis and Emergency Risk Communication (CERC) model emphasizes "to communicate in accurate, credible, timely, and reassuring" (p.45) ways during high-uncertainty situations (Reynolds & Seeger, 2005). However, CERC was designed for terrestrial crises with certain established response protocols. Extraterrestrial intelligence discovery presents a crisis without precedent, requiring novel theoretical integration.

Weick's theory of enacted sensemaking (1988) provides a critical insight into how collective understanding forms during unprecedented events. In moments of profound uncertainty, initial actions by authorities create the very environment that people are trying to comprehend. For ISO discoveries, early scientific statements will not merely describe reality but will construct it, shaping public perception. Moreover, Weick argues that strong commitment to early interpretations during crisis situations can create "blind spots" that prevent decision-makers from recognizing critical information or alternative explanations, potentially allowing crises to escalate.

Complementing sensemaking theory, Coombs' Situational Crisis Communication Theory (SCCT) offers an evidence-based model for managing reputational threat based on attributed responsibility levels (Coombs, 2007). For extraterrestrial intelligence discovery, the scientific community must proactively manage its perceived responsibility for information clarity and timeliness to maintain credibility throughout extended verification periods.

SCCT's theory foundation becomes particularly relevant for ISO discoveries. The theory identifies three crisis clusters: (1) *victim*, where the organization has minimal responsibility, (2) *accidental*, moderate responsibility, and (3) *preventable*, high responsibility. ISO discoveries uniquely span all three clusters simultaneously. Scientists can be victims of incomplete data, accidentally misinterpret observations, or be perceived as preventable withholding information. Such a multi-cluster nature requires adaptive response strategies that acknowledge varying levels of institutional responsibility as evidence evolves. For instance, initial detection (victim cluster) requires expressing concern and information sharing, while misinterpretation (accidental cluster) demands justification and corrective action, and perceived withholding (preventable cluster) necessitates apology and comprehensive disclosure.

**Risk Amplification and Information Dynamics**

The Social Amplification of Risk Framework (SARF) reveals how risk signals undergo transformation through social stations including media, opinion leaders, and social networks (Kasperson et al., 1988). For ISO discoveries, these amplification processes can transform measured scientific announcements into social phenomena far exceeding their initial scope, generating ripple effects across economic, religious, and political domains. The growth of social media has transformed this landscape, allowing crises to spread faster than on traditional platforms and making them more difficult to manage due to increased interactivity and a "vast spectrum for public opinion" (Apuke & Tunca, 2018).

The SARF framework's relevance to ISO communication extends beyond traditional risk scenarios. Unlike conventional hazards with bounded consequences, extraterrestrial intelligence discovery carries unbounded psychological and social implications. Kasperson et al.'s (1988) concept of "signal value", the degree to which an event provides new information about the magnitude of risk, reaches maximum levels for ISO discoveries. Every piece of evidence, regardless of actual significance, can trigger disproportionate social responses. Digital amplification stations now include algorithm-driven social media platforms, conspiracy theory networks, and sponsored disinformation channels, each capable of transforming scientific uncertainty into social certainty within hours.

The principles of effective public health risk and crisis communication (Covello, 2003) provide essential operational guidance. A foundational concept is the necessity of building trust and credibility, which is not expected but earned through deliberate, ethical actions. This process begins by accepting and involving all stakeholders as legitimate partners early in the decision-making process, revealing respect for those affected. Effective communication requires being truthful, honest, and open, especially when information is incomplete or evolving. Rather than over-reassuring the public, the guide suggests a transparent approach that includes sharing information as soon as possible, discussing data uncertainties, and even identifying worst-case estimates to avoid perceptions of withholding information. Furthermore, messages must be delivered with clarity and compassion, using clear, non-technical language tailored to the audience and acknowledging the validity of their emotions. The guidelines also highlight the critical importance of careful and thorough planning, which involves setting clear communication objectives, pre-testing messages, and training spokespersons. Such a comprehensive approach underscores that effective crisis communication is a planned and sustained effort built on transparency and partnership.

**Learning from Communication Cases**

***COVID-19: The Cost of Uncertainty Mismanagement***

The COVID-19 pandemic exemplifies failed uncertainty communication at global scale. Initial inconsistent messaging regarding transmission mechanisms, mask efficacy, and threat severity created an information vacuum rapidly filled by an "infodemic" of misinformation (Zarocostas, 2020). The World Health Organization's January 14, 2020, tweet stating "no clear evidence of human-to-human transmission" (World Health Organization, 2021) technically reflected available evidence but failed to convey uncertainty levels, undermining credibility when transmission was later confirmed (Sharma et al., 2021). Health authorities' reluctance to acknowledge evolving understanding eroded public trust when guidance changed. For ISO discoveries, where evidence will emerge gradually and interpretations will evolve, transparent uncertainty communication becomes essential to maintain credibility.

Key lessons from COVID-19 include the role of normalizing scientific uncertainty in public health communication. Research showed that acknowledging uncertainty while reinforcing its expected nature can reduce ambiguity-averse responses, including heightened risk perceptions and emotional distress (Han et al., 2021).

New Zealand's successful approach, characterized by clear uncertainty acknowledgment, consistent cross-party messaging, and regular empathetic communication, maintained public trust despite policy changes (Beattie & Priestley, 2021). Critical to New Zealand's success was establishing a unified communication strategy through the Unite Against COVID-19 campaign, which provided coordinated messaging across multiple channels and was informed by evidence-based research from experts including psychologist Sarb Johal (Beattie & Priestley, 2021). The campaign's central platform, the covid19.govt.nz website, served as a continuously updated single source of truth that competed effectively with social media misinformation. For ISO discoveries, a similar "Global ISO Status Portal" maintained by a designated international astronomical authority can provide real-time observational data, expert interpretations, and uncertainty metrics, preventing information vacuums that speculation and misinformation inevitably fill.

### Fukushima: Technical Complexity and Trust Collapse

The 2011 Fukushima Daiichi nuclear disaster illustrates how technical complexity combined with institutional self-protection destroys trust (Funabashi & Kitazawa, 2012; Perko, 2015). Delayed radiation data release, overly technical language that obscured rather than clarified risks, and visible disconnects between government and corporate statements created perception of coverup (Funabashi & Kitazawa, 2012; Perrow, 2011). Less than 16% of media articles contained radiation units, and when used, inconsistent measurements (mSv, μSv, Bq/kg) created confusion (Perko et al., 2015). Tokyo Electric Power Company's repeated use of "unanticipated" for historically precedented tsunami heights became emblematic of obfuscation. The use of specialized terminology without translation for public understanding generated confusion that conspiracy theories exploited. The failure to use SPEEDI evacuation data and dismissal of officials who mentioned "meltdown" compounded distrust (Funabashi & Kitazawa, 2012).

For ISO communication, Fukushima teaches that technical accuracy must be balanced with public accessibility. Scientific precision cannot come at the cost of comprehension. Message architecture must layer complexity, providing immediate understanding for general audiences while offering technical depth for those seeking it. The delayed admission of core meltdowns, acknowledged only after two months despite internal awareness within hours, reveals how institutional credibility, once lost, becomes nearly impossible to restore (Funabashi & Kitazawa, 2012).

### Asteroid 99942 Apophis: Managing Graduated Threat Levels

The 2004 discovery of asteroid Apophis, which reached Level 4 on the Torino Scale (Binzel, 2000; Reddy et al., 2022), provides the closest precedent to graduated ISO threat assessment. Initial calculations indicating a 2.6% Earth impact probability in 2029 (Souchay et al., 2014) required careful calibration between generating appropriate concern and preventing panic. NASA's communication strategy employed graduated disclosure synchronized with observational refinement, providing regular updates as orbital calculations improved. By gradually reducing threat level as observations refined orbital calculations, NASA revealed the viability of graduated disclosure protocols. NASA's announcement balanced urgency with measured tone.

NASA's specific messaging architecture for Apophis revealed graduated disclosure effectiveness. Initial announcement (December 23, 2004): "NASA scientists are tracking an asteroid with a 1-in-300 chance of impacting Earth in 2029" (Yeomans et al., 2004a). This factual statement avoided sensationalism while acknowledging uncertainty. December 24 update: "Nevertheless, the odds against impact are still high, about 60 to 1, meaning that there is a better than 98% chance that new data in the coming days, weeks, and months will rule out any possibility of impact in 2029" (Yeomans et al., 2004a). December 28 resolution: "Additional observations have ruled out 2029 impact" (Yeomans et al., 2004b). With each update, the agency utilized visualization tools showing probability distributions and confidence intervals, helping media and public psychologically adjust to uncertainty reduction over time.

The Apophis case reveals that people can process evolving risk assessments when they are provided clear information for understanding uncertainty reduction. The Torino Scale's 0-10 classification system (Binzel, 2000), like the proposed Loeb Scale (Loeb, 2025a), provide consistent communication across institutions while helping media convey appropriate threat levels without sensationalism.

While these historical cases, COVID-19, Fukushima, and Apophis, reveal the role of *uncertainty management, technical accessibility*, and *graduated disclosure*, they also reveal a critical gap: the absence of a unified framework that addresses how diverse populations process paradigm-shifting information through different cognitive and emotional channels. The Apophis case showed that clear classification systems and consistent messaging can work, but primarily reached audiences already attuned to scientific communication. For the unprecedented challenge of potential extraterrestrial intelligence discovery, which would affect all of humanity regardless of scientific literacy or cultural context, a more comprehensive approach is required.

**The SPECtrum of Rhetoric Intelligences Framework**

The SPECtrum of Rhetoric Intelligences theory (SPEC/RI; Teer, 2020), developed through analysis of effective political communication, identifies four distinct cognitive channels through which individuals process information. This framework proves essential for anomalous ISO communication to the public, as different populations will process paradigm-shifting information through distinct cognitive and emotional filters. SPEC/RI recognizes that individuals dynamically engage different processing modes which consist of context, stress level, and information type.

*Systematic Intelligence Quantification (SQ)*

Systematic Intelligence resonates with those seeking authority and structure. These individuals trust institutional validation, peer review, and scientific consensus. For ISO discoveries, SQ messaging emphasizes international cooperation, methodological rigor, and convergent evidence from multiple independent sources. Messages targeting SQ must maintain clear information hierarchy and cite authoritative sources consistently. Example: "The International Astronomical Union confirms that fourteen independent research teams have verified the anomalous characteristics of interstellar object 3I/ATLAS through peer-reviewed analysis".

If ISO communications initially overemphasize SQ channels, using technical language and focusing on orbital mechanics will reach scientific audiences effectively but will fail to engage broader publics who process information differently.

*Practical Intelligence Quantification (PQ)*

Practical Intelligence focuses on concrete outcomes and personal relevance. PQ-oriented individuals may ask: "What does this mean for my daily life?" Effective PQ messaging for ISO discoveries must address immediate concerns about safety, economic impacts, and continuation of normal activities. These messages translate cosmic significance into comprehensible, actionable information. Example: "The object poses no threat to Earth. All essential services, from healthcare to transportation, continue operating normally."

In case of an absence of PQ, messaging will lead to unnecessary anxiety among people. Clear statements that the object poses no threat and that daily life would continue unchanged can prevent speculation about immediate danger.

### Emotional Intelligence Quantification (EQ)

Emotional Intelligence connects through narrative and human experience. Stories of scientists making discoveries, historical parallels to previous paradigm shifts, and emphasis on humanity's shared journey prove essential. EQ messaging acknowledges feelings of awe, uncertainty, and excitement while providing reassurance through human connection. Example: "Dr. Smith, who first noticed the anomaly, describes feeling 'humbled by the possibility we might not be alone, it reminds us that we're all part of something larger'".

Effective ISOs coverage could include profiles of discovering astronomers and their personal reactions. These human elements can create abstract concepts accessible and help publics connect emotionally with the discovery process.

### Creative Intelligence Quantification (CQ)

Creative Intelligence captures attention through unexpected approaches and novel presentations. Visual representations, analogies, artistic collaborations, and even appropriate humor make complex concepts accessible. CQ elements ensure message retention and viral spread through social media channels. Example: Interactive 3D models allowing users to explore the object's trajectory or comparing its shape to familiar objects like "a cosmic pancake."

Interactive visualizations of ISOs trajectory and artist conceptions of possible configurations reveal CQ's power to engage audiences who may otherwise ignore technical announcements. Such creative elements generated discussion and sharing far exceeding traditional scientific communications.

While the SPEC/RI framework provides essential channels for multi-audience communication, the contemporary information ecosystem introduces an unprecedented variable that fundamentally alters crisis communication dynamics. Unlike previous paradigm-shifting discoveries, any potential ISO announcement will be mediated not only through traditional channels but through billions of AI-powered interactions that can amplify, distort, or reframe scientific findings within seconds of release.

**Artificial Intelligence in Crisis Communication: Opportunities and Threats**

AI assistants will likely serve as primary interpreters of ISO discoveries for the public, creating unprecedented challenges. The first threat involves human vulnerability to synthetic media, with detection accuracy barely above chance levels at approximately 51% (Cooke et al., 2025). Within hours of any ISO announcement, fabricated videos of scientists making false claims or synthetic "alien messages" could circulate globally. Similar patterns emerged during COVID-19, where information from questionable and reliable sources spread at similar rates, though the volume of misinformation varied by platform (Cinelli et al., 2020), and ISO discoveries would face similar amplification dynamics.

The second threat concerns AI response accuracy. Although top models achieve hallucination rates below 5% in controlled conditions (Vectara, 2025), they reach 83% error rates under adversarial conditions (Omar et al., 2025). Millions of users asking AI assistants about ISO discoveries could receive confidently stated but incorrect interpretations, transforming scientific uncertainty into false certainty. These AI systems, trained on science fiction alongside scientific literature, may generate plausible but entirely fabricated explanations when faced with unprecedented scenarios.

Without addressing both human susceptibility to deepfakes and AI tendency to hallucinate, humanity's potentially most significant discovery risks being overwhelmed by synthetic deception and algorithmic misinformation. These AI-specific challenges and protocols must be integrated into comprehensive communication framework for ISO discoveries. The convergence of human cognitive diversity (as captured by SPEC/RI) and AI mediation creates unprecedented complexity that traditional crisis communication models cannot adequately address. We now turn to synthesizing these theoretical insights and practical challenges into a unified framework.

### Theoretical Framework Development

Having examined both the lessons from analogous communication cases and the multi-channel processing model of SPEC/RI (Teer, 2020), we now describe how these features are synthesized into the Adaptive Communication Framework (ACF). The ACF development follows an established theory-building approach in management and communication sciences (Jaakkola, 2020; Whetten, 1989), integrating three analytical components:

First, we reviewed foundational crisis communication theories, including CERC, sensemaking theory, SCCT, and SARF, to identify core principles applicable to unprecedented discovery scenarios. The theoretical review focused on mechanisms of uncertainty communication, trust maintenance, and information amplification dynamics.

Second, we analyzed three analogous communication cases that share critical characteristics with potential ISO discoveries: COVID-19 pandemic (evolving scientific understanding), Fukushima nuclear disaster (technical complexity and trust dynamics), and asteroid Apophis (graduated astronomical threat). For each case, we examined official communications, documented public responses, and identified successful strategies and critical failure points.

Third, we integrated the SPEC/RI framework to address multi-channel communication needs, recognizing that different audiences process paradigm-shifting information through distinct cognitive and emotional filters. We mapped each intelligence type to specific ISO communication challenges and developed corresponding message strategies.

Our approach recognizes that effective communication of potentially paradigm-shifting discoveries requires more than information transfer. It demands careful orchestration of multiple cognitive, emotional and social channels to facilitate collective sensemaking and collective efficacy during periods of unprecedented uncertainty. This orchestration must occur simultaneously across channels to prevent any single narrative from dominating public perception. Such a theoretical synthesis produced the Adaptive Communication Framework (ACF, see Figure 1), which provides both analytical consideration of communication dynamics and prescriptive guidance for ISO discovery announcements.

**Figure 1**

*The Adaptive Communication Framework (ACF) for ISO discovery*

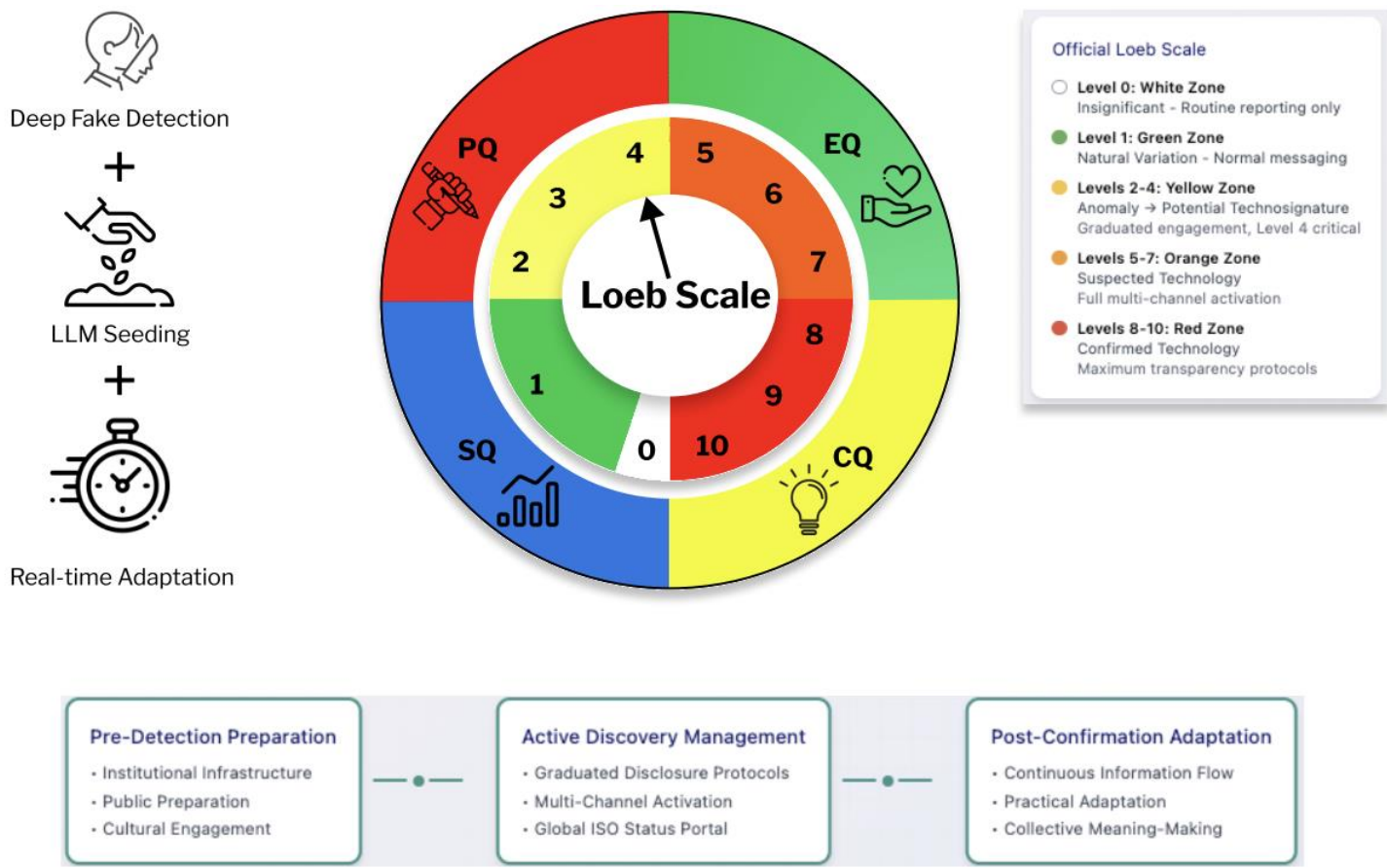

*Note.* The Adaptive Communication Framework operationalizes multi-channel crisis communication for potential extraterrestrial intelligence discoveries. The central Loeb Scale (0-10) determines communication intensity, with color gradients indicating escalating significance: white (0, insignificant), green (1, natural), yellow (2-4, anomalous), orange (5-7, suspected technology), and red (8-10, confirmed technology). Four SPEC/RI intelligence channels (Systematic/SQ, Practical/PQ, Emotional/EQ, Creative/CQ) surround the scale, activated simultaneously at Level 4+ to prevent narrative fragmentation. Left-side AI safeguards (deep fake detection, LLM seeding, real-time adaptation) address algorithmic mediation of public

understanding. The three-phase operational timeline (Pre-Detection Preparation → Active Discovery Management → Post-Confirmation Adaptation) includes bidirectional feedback loops for dynamic response adjustment. This integrated architecture ensures coordinated global communication across diverse cognitive processing styles while mitigating AI-era information threats.

## The Adaptive Communication Framework

### Phase 1: Pre-Detection Preparation

Effective ISO communication requires infrastructure established before discovery pressures create competitive dynamics. The preparation phase focuses on establishing authoritative structures, building public literacy, and developing cultural frameworks.

**Institutional Infrastructure:** The foundation involves establishing an International Committee for Extraterrestrial Communication (ICEC) under a designated international astronomical authority, paralleling the International Atomic Energy Agency's role in nuclear incidents. This ICEC role is to develop standardized protocols, credential legitimate spokespersons, and maintain rapid response capabilities for rapid activation. Committee membership must include not only scientists, AI specialists, political leaders and policy makers but also communication specialists, social psychology scientists, religious leaders, and ethicists to ensure multiple perspectives inform messaging strategies.

**Public Preparation:** Comprehensive FAQ databases must cover economic implications, security concerns, and daily life impacts, translated into major languages and adapted for cultural contexts. These resources provide immediate, authoritative answers that prevent speculation from filling information voids during actual discoveries. Educational initiatives must normalize

ISO detection within broader astronomical education, emphasizing scientific process rather than sensational possibilities.

Cultural Engagement: Long-term public engagement through storytelling creates psychological resilience. Documentary content exploring humanity's cosmic questions, profiles of scientists conducting research, and educational materials that acknowledge both excitement and uncertainty prepare public psychology for potential discoveries. Such narratives must emphasize humanity's collective journey rather than competition or conflict frames that could generate destructive dynamics.

**Phase 2: Active Discovery Management**

When potential technosignature evidence emerges, communication must balance scientific uncertainty with public information needs. The framework provides level-specific protocols calibrated to the Loeb Scale (see Loeb, 2025a):

**Levels 0-3 (Natural to Anomalous):** Communication remains primarily within scientific channels. Public messaging emphasizes the discovery process rather than premature conclusions. Systematic intelligence channels dominate with authoritative updates from established institutions. Practical messaging addresses the reasons for the observation's necessity. Emotional content shares the scientists' curiosity and rigorous methodology. Creative elements may include real-time data visualizations allowing public participation in the analysis.

**Level 4 (Potential Technosignature Indicators):** This threshold triggers enhanced communication protocols. All four intelligence channels are activated simultaneously with coordinated messaging: (1) *Systematic*: "International teams are conducting coordinated observations following established protocols", (2) *Practical*: "No immediate action required; the object poses no threat", (3) *Emotional*: "Scientists worldwide are working together, embodying humanity's collective curiosity", and (4) *Creative*: Interactive trajectory models and comparison graphics engage public attention constructively.

Building on Loeb's (2025b) operational principles for responding to potential alien technology, Level 4 activation includes establishing a centralized "Global ISO Status Portal", a single, authoritative, real-time information source accessible to all humanity. This portal aggregates all observational data, expert analyses, and uncertainty assessments in one location, preventing the information fragmentation that enables misinformation. As Loeb emphasizes, information must be shared in full among all humans, since the response affects our shared future. Features include real-time telescope feeds, expert commentary with uncertainty bounds, predictive trajectory modeling, and multi-language accessibility, ensuring no nation gains information advantage while maintaining complete transparency.

Level 4 activation simultaneously triggers AI Communication Protocol. The ICEC transmits standardized information packages to all major AI platforms, ensuring consistent responses to inevitable user queries. Automated systems must begin scanning for deepfakes across platforms, flagging suspicious content for review within minutes of upload. Social media companies receive formal requests to modify recommendation algorithms, prioritizing ICEC-verified content over unverified claims.

The Global ISO Status Portal incorporates AI capabilities for personalized information delivery based on users' SPEC/RI profiles and cultural contexts. Natural language processing enables conversational queries with blockchain-authenticated responses. This AI-enhanced infrastructure addresses a critical gap: while traditional media can be engaged through established channels, AI systems operate through opaque neural networks requiring proactive information seeding. Without pre-positioned, verified content, language models may generate unpredictable interpretations which rely on science fiction training data rather than scientific evidence, potentially fragmenting public understanding when unity is essential.

**Levels 5-7 (Suspected Technology)**: Communication intensity escalates with predetermined triggers. Messages acknowledge increasing confidence while maintaining scientific caution. Practical intelligence becomes paramount with clear statements about threat assessment and response protocols. Emotional messaging pivots toward unity themes emphasizing humanity's shared stake in the discovery. Creative approaches may include collaborative global events or cultural celebrations of cosmic citizenship.

**Levels 8-10 (Confirmed Technology)**: Full transparency protocols activate with immediate, simultaneous global announcement through all available channels. The message architecture follows a specific two-minute sequence repeated across multiple messengers and platforms: (1) *systematic establishment of facts* (30 seconds), (2) *practical immediate implications* (30 seconds), (3) *emotional human context* (30 seconds), and (4) *creative future vision* (30 seconds). This ensures core information reaches diverse audiences before speculation and misinformation can dominate narrative formation.

**Phase 3: Post-Confirmation Adaptation**

Following technosignature confirmation, humanity faces unprecedented coordination challenges. As Loeb (2025b) notes, decisions must be prompt because of the physical proximity of the visiting devices, markedly different from traditional SETI protocols assuming distant radio sources. The ACF's post-confirmation phase operationalizes Loeb's principles through three parallel tracks:

**Continuous Information Flow**: Systematic channels provide updates every six hours, even if stating "no change", preventing speculation-filled voids. Scientific panels explain implications across disciplines, biology (implications for life's universality), physics (novel propulsion mechanisms), philosophy (humanity's cosmic significance), and theology (integration with religious frameworks). Following Loeb's tenth principle (Loeb, 2025b), governments maintain "continuous contact, exchanging real-time alerts" through the ICEC's secure communication network.

**Practical Adaptation Infrastructure**: Economic stabilization protocols must be activated, with central banks coordinating to prevent market panic while addressing the need to redirect "a substantial fraction of the 2.4 trillion dollars allocated annually to military budgets" (Loeb, 2025b) toward new defense technologies and scientific research. Mental health support systems, specifically trained for existential discovery counseling, deploy through existing healthcare infrastructure.

**Collective Meaning-Making**: Humanity requires practical frameworks for processing transformed cosmic status. Historical parallels (Copernican revolution, discovery of deep time, DNA structure) provide cognitive anchors that help humanity process paradigm-shifting discoveries by anchoring the unfamiliar in the familiar (Ghilani et al., 2017). Global commemoration events, simultaneous observations, cultural celebrations of cosmic citizenship, artistic collaborations, create shared experience transcending national boundaries. As Loeb emphasizes: "we are all in the same boat, and if one of us rocks the boat, all of us might be at risk of drowning" (Loeb, 2025b).

## Discussion

The Adaptive Communication Framework (ACF) addresses critical challenges in managing humanity's potential encounter with evidence of extraterrestrial intelligence. Our theoretical analysis reveals several key insights and implementation challenges.

**Managing Uncertainty**

Unlike binary crisis announcements, ISO evidence emerges through increasing confidence levels rather than definitive moments. The framework's graduated approach allows communication to evolve with evidential certainty, preventing premature conclusions while maintaining public engagement throughout extended verification periods. The conceptual and practical framework must balance scientific rigor with public information needs, acknowledging that withholding information too long risks losing narrative control to speculation and misinformation.

**Cultural and Religious Integration**

Extraterrestrial intelligence confirmation intersects with fundamental belief systems differentially across cultures. While some audiences might primarily process discoveries through scientific channels, other populations may prioritize religious or philosophical interpretations. The framework's multi-channel allows culturally adapted messaging without compromising core information. Religious authorities must be engaged as partners rather than obstacles, recognizing that spiritual frameworks for understanding cosmic intelligence may provide psychological resilience for billions of people.

**International Coordination Challenges**

ISO discovery transcends national boundaries, requiring unprecedented international cooperation. While the proposed ICEC provides coordination mechanisms, practical implementation faces significant obstacles. Nations may attempt to implement information control for perceived advantage, institutions may compete for announcement priority, and cultural differences may generate conflicting messages. The framework must maintain core message consistency while allowing regional adaptation. Success requires pre-negotiated agreements establishing communication protocols before discovery pressure creates competitive dynamics.

## Misinformation Mitigation

The digital information ecosystem guarantees that ISO-related misinformation will spread rapidly, probably outpacing official channels. The framework's pre-positioned infrastructure and platform partnerships provide some defense, but complete prevention remains impossible. Strategy must focus on rapid response and credibility maintenance rather than information control. By activating all four intelligence channels simultaneously, accurate information reaches audiences through their preferred cognitive paths, reducing susceptibility to alternative narratives. Platform partnerships with major technology companies must ensure official information receives algorithmic priority during critical announcement periods.

The conceptual-practical framework's success depends critically on coordinating thousands of independent AI systems during the crucial first hours following any official announcement. Unlike traditional media, AI outputs cannot be perfectly predicted or controlled. Even with pre-positioned information, language models may generate unexpected interpretations based on implicit training biases about extraterrestrial intelligence. Furthermore, adversarial actors could deploy AI-generated disinformation campaigns, creating thousands of plausible but false narratives targeting specific populations through personalized misinformation. These challenges require not only defensive measures but offensive capabilities to counter AI-generated disinformation with equally sophisticated corrective messaging, transforming the ACF into an active information warfare system rather than passive communication protocol.

**Conclusions**

As humanity enters an era of routine interstellar object detection, the probability of identifying objects with characteristics suggesting artificial origin is elevating. The Adaptive Communication Framework (ACF) provides essential theoretical and practical infrastructure for managing this potentially transformative discovery, integrating crisis communication theory with multi-channel messaging strategies to address both information dissemination and psychological adaptation needs.

The urgency of implementation cannot be overstated. Currently, humanity is not prepared for a potential threat from alien technology (Loeb, 2025b). The ACF provides the communication infrastructure necessary to transform this vulnerability into preparedness. The proposed conceptual framework establishes pre-discovery communication architecture that determines whether humanity responds with coordinated wisdom or chaotic confusion.

Immediate implementation requires five critical actions. First, establish the International Committee for Extraterrestrial Communication (ICEC), under a designated international astronomical authority. Second, develop comprehensive message templates addressing all four intelligence types for each Loeb Scale level. Third, launch public preparation programs integrating ISO education into existing astronomical outreach, beginning immediately with existing resources. Fourth, negotiate platform partnerships with major technology companies ensuring official information priority during announcement periods, with agreements finalized. Fifth, conduct annual simulation exercises, involving all stakeholder groups to refine protocols and build institutional preparedness.

The framework's strength lies not in preventing all negative outcomes but in providing structured approaches for managing inevitable complexity. By acknowledging that different individuals process paradigm-shifting information through distinct cognitive and emotional filters, the ACF ensures messages reach diverse global audiences effectively. The convergence of technological capability, public readiness, and cosmic probability makes comprehensive communication preparation essential rather than optional.

Whether the next interstellar visitor proves natural or artificial, billions of people will learn about it through AI assistants whose responses, accurate or hallucinatory, calming or inflammatory, will shape humanity's understanding. The Adaptive Communication Framework (ACF) ensures both human and artificial intelligence systems work in concert rather than opposition. Without immediate implementation including AI safeguards, humanity's most profound potential discovery risks unfolding through algorithmic chaos rather than coordinated communication. The convergence of enhanced detection capability, AI ubiquity, and absent protocols creates unprecedented vulnerability. The tools for discovery are ready; our communication infrastructure, both human and artificial, must be equally prepared when discovery arrives.